\documentclass[a4paper,11pt]{article}
\usepackage{amsmath, amsthm, amsfonts, amssymb}
\usepackage[usenames,dvipsnames]{xcolor}
\colorlet{color1}{NavyBlue}
\usepackage[colorlinks=true,allcolors=color1]{hyperref}
\usepackage{amssymb}
\usepackage{physics}
\usepackage{mathrsfs}
\usepackage{dsfont}
\usepackage{graphicx}
\usepackage{xfrac,overpic}
\usepackage[utf8]{inputenc}

\begin{document}

\title{On destabilising quasi-normal modes with a radially concentrated perturbation}

\author{Valentin Boyanov \vspace*{.2cm}\\ \textit{CENTRA, Departamento de F\'{\i}sica, Instituto Superior T\'ecnico -- IST,}\\ \textit{Universidade de Lisboa -- UL, Avenida Rovisco Pais 1, 1049 Lisboa, Portugal}}

\maketitle

\begin{abstract}
	In this work we explore some aspects of the spectral instability of back hole quasi-normal modes, using a specific model as an example. The model is that of a small bump perturbation to the effective potential of linear axial gravitational waves on a Schwarzschild background, and our focus is on three different aspects of the instability: identifying and distinguishing between the two different types of instabilities studied previously in the literature, quantifying the size of the perturbations applied to the system and testing the validity of the pseudospectral numerical method in providing a convergent result for this measure, and finally, relating the size and other features of the perturbation to the degree of destabilisation of the spectrum.
\end{abstract}

\section{Introduction}

The quasi-normal modes (QNMs) of black holes (BHs) have been shown to suffer from a spectral instability, which shifts these characteristic frequencies by disproportionately large distances in the complex plane when the system is subjected to seemingly small environmental perturbations. This has been shown through calculations of the QNM spectrum after a variety of generic perturbations are added to the system~\cite{JaramilloPRX,Jaramillo2021,Sheikh2022,Nollert1996,NollertPrice,Daghigh2020,Qian2020,Boyanov2022,Arean,Warnick,dS,Destounis2023,Boyanov2023,Ianniccari2024,Gogoi2024,Skvortsova2024}, and quantitatively explored through the full pseudospectrum of the linear perturbation problem~\cite{JaramilloPRX,Sheikh2022,Boyanov2022,Arean,Warnick,dS,Destounis2023,Boyanov2023,Destounis2021,GaussB}, generally in a physically motivated norm~\cite{Gasperin2021}. On the other hand, a seemingly qualitatively different instability has been observed when the perturbation involved is specifically the addition of a single small ``bump" to the effective potential of the propagating waves at different distances from the black hole horizon, intended to mimic some radially concentrated distribution of matter~\cite{elephant,Berti2022,Cardoso2024}, or, more generally, the addition of a second length scale in the problem~\cite{Cardoso2024,Qian2024}. The former of these approaches stands out through its consistent attempt to precisely quantify the magnitude of the perturbations applied to the system, and thus the amount by which the QNM migration exceeds the threshold of stability. The latter approach, on the other hand, has found a rich phenomenology which includes the appearance of new branches of QNMs which can contain modes with a longer lifetime than the BH fundamental mode, akin to the ``shape resonances" discussed in e.g.~\cite{Zworski,BindelZworski}.

However, in spite of the varied nature of these results, the endeavour to obtain a complete physical picture of this instability has not yet come to fruition. On the one hand, not all results have been put in the context of the quantitative scheme devised in \cite{JaramilloPRX} involving the energy norm. On the other hand, this scheme itself may not be the most adequate for precise quantitative conclusions. As discussed already in \cite{JaramilloPRX}, two perturbations of the same energy norm can have vastly different destabilisation effects depending on their high-wave-number content (i.e. on the magnitude of derivatives in $r$ of the perturbation function). Additionally, as shown in \cite{Boyanov2023} for one particular model, some of the numerically computed quantities involved in pseudospectrum calculations may not be well behaved in the continuum limit.

The present work is intended as a short review of the subject, particularly highlighting some of the above mentioned issues, using a specific example to further clarify them and progress towards their resolution. The example system chosen is that of axial gravitational perturbations on a Schwarzschild BH, with a gaussian bump added to the effective potential of their governing wave equation, akin to the one used in \cite{elephant}.

Section \ref{s2} provides a brief overview of the QNM instability and the tools used to capture and quantify it. Section \ref{s3} uses the gaussian bump setup for: \ref{s31} providing a simple example of the instability, \ref{s32} bringing the results of the analysis in \cite{elephant} to the context of the energy norm, quantifying the ``smallness" of the bumps added to the potential, as well as discussing the emergence of new mode branches due to a qualitative change in the phase space of the evolution operator~\cite{Zworski,BindelZworski} (see in particular footnote 9 of~\cite{Jaramillo2022} and footnote 13 of~\cite{Gasperin2021}), \ref{s33} exploring the dependence of the degree of destabilisation on the ``high-wave-number" content~\cite{JaramilloPRX}, or sharpness, of the added bump, as well as presenting an analysis regarding the numerical convergence of the results. Finally, section \ref{s4} presents a summary of the conclusions which can be drawn from this analysis and used as guidance for future work in this field.

\section{Linear perturbations and norm}\label{s2}

The background spacetime we will work with is the Schwarzschild geometry,
\begin{equation}
ds^2=-f(r)dt^2+\frac{1}{f(r)}dr^2+r^2d\Omega^2,
\end{equation}
where the redshift function reads $f(r)=1-2M/r$, and $d\Omega^2$ is the line element of the unit sphere. The maximal extension of this spacetime has a bifurcate Killing horizon at $r=2M$, though for QNMs the important part is the outgoing horizon which in the future is equivalent to the event horizon of a dynamically formed (non-evaporating) black hole.

The dynamics of linear perturbations around this background is given by a wave equation,
\begin{equation}\label{waveeq}
-\partial_t^2\phi+\partial_{r^*}^2\phi-V(r)\phi=0,
\end{equation}
where $r^*$ is the tortoise coordinate, $dr^*=dr/f(r)$, and the potential $V$ depends on the nature of the perturbation and on its angular multipole number $\ell$. In the example below we will analyse the case of axial gravitational perturbations,
\begin{equation}\label{potential}
V=\frac{f}{r^2}\left[\ell(\ell+1)-\frac{6M}{r}\right].
\end{equation}
Quasi-normal modes are a discrete set of analytic solutions to \eqref{waveeq} which behave as ingoing waves,
\begin{equation*}
	\phi\sim e^{i\omega(t+r^*)}
\end{equation*}
at the horizon, and as outgoing waves,
\begin{equation*}
	\phi\sim e^{i\omega(t-r^*)}
\end{equation*}
at infinity. These conditions can be imposed geometrically in the wave equation by expressing it in a hyperboloidal coordinate system~\cite{Zenginoglu2011,PanossoMacedo2024}, with the transformation $\{t,r^*\}\to\{\tau,\chi\}$ given by
\begin{equation}
\begin{split}
\frac{t}{2M}&=\tau-h(\chi),\\
\frac{r^*}{2M}&=g(\chi),
\end{split}
\end{equation}
where $h(\chi)\sim g(\chi)$ when approaching the horizon, and $h(\chi)\sim -g(\chi)$ when approaching infinity. A standard choice is the so-called minimal gauge~\cite{PanossoMacedo2023}, which for the Schwarzschild case is given by
\begin{equation}
\begin{split}
h(\chi)&=\log(1-\chi)-\frac{1}{\chi}-\log\chi,\\
g(\chi)&=\log(1-\chi)+\frac{1}{\chi}+\log\chi.
\end{split}
\end{equation}
The compactified radial coordinate $\chi=2M/r$ spans the range $\chi\in(0,1)$ between (future null) infinity and the (future) horizon. The QNM boundary conditions now amount to simply requiring regularity of the solutions at the boundaries.

Following ref.~\cite{JaramilloPRX}, we perform this coordinate transformation along with an order reduction in time through the introduction of the auxiliary variable $\psi=\partial_\tau\phi$, recasting the problem in the form
\begin{equation}\label{evol}
iL\,u=\partial_\tau u,
\end{equation}
where
\begin{equation}
u=\begin{pmatrix}
\phi\\ \psi
\end{pmatrix}, \qquad L=\frac{1}{i}\begin{pmatrix}
0&\mathbb{I}\\L_1&L_2
\end{pmatrix},
\end{equation}
with
\begin{equation}
\begin{split}
	L_1&=\frac{p}{w}\partial_\chi^2+\frac{p'}{w}\partial_\chi-\frac{q}{w},\\
	L_2&=2\frac{\gamma}{w}\partial_\chi+\frac{\gamma'}{w},
\end{split}
\end{equation}
and we have defined the functions
\begin{equation}
w=\frac{|g'|}{g'^2-h'^2},\quad p=\frac{1}{|g'|},\quad \gamma=\frac{h'}{|g'|},\quad q=|g'|V,
\end{equation}
a prime denoting differentiation with respect to $\chi$. The QNM frequency spectrum can be defined~\cite{Warnick2013,Ansorg2016} as the eigenvalues of the evolution operator $L$, or equivalently as the poles of the resolvent operator
\begin{equation}
R_L(\lambda)=(L-\lambda\mathbb{I})^{-1}.
\end{equation}
Since $L$ is non-self-adjoint (due to the dissipative boundaries of the problem), solutions to the wave equation cannot be expressed simply as convergent series of the eigenfunctions, i.e. of QNMs. Additionally, and crucially, the QNM frequencies can be unstable to ``small" perturbations of the system. Perturbations can come in many shapes and sizes, and the effect they can have on the spectrum is just as varied. The instability originally studied in ref.~\cite{JaramilloPRX} consists in the displacement of modes in the complex plane by distances much larger than the size (energy norm) of the perturbations would allow for a spectrally stable operator.

However, one interesting conclusion in ref.~\cite{JaramilloPRX} is the apparent stability of the fundamental mode, and the absence of any displaced overtones which would have a slower decay rate (smaller imaginary part) than this fundamental one after a perturbation. In contrast to this result, ref.~\cite{elephant} found that perturbing the effective potential with a seemingly very small bump placed sufficiently far from the horizon can easily destabilise the fundamental mode, leaving a mode with a much smaller imaginary part as the new fundamental one. The apparent contradiction between these conclusions is mainly due to a qualitative difference in the type of perturbations and instability considered. We will now present a summary of some aspects of these two analyses, and highlight the differences between them. Then, in the following section, we will proceed to analyse an example, originally treated in ref.~\cite{elephant}, which turns out to lead to a combination of both destabilising effects.

\subsection{Mode displacement and pseudospectrum}

The case of QNM instability analysed in~\cite{JaramilloPRX} and related works is one in which (at least part of) the already existing BH QNM spectrum is displaced by a disproportionately large amount due to a small perturbation to the operator $L$. The smallness of this perturbation is defined quantitatively through the energy norm~\cite{Gasperin2021}, which has a natural physical interpretation. The overall instability to \textit{any} perturbation of $L$ is captured by the pseudospectrum in this norm, which is defined as
\begin{equation}\label{pseudore}
\sigma^\epsilon(L)=\{\lambda\in\mathbb{C}:\|R_L(\lambda)\|_E>1/\epsilon\}, 
\end{equation}
where $\|\cdot\|_E$ indicates the energy norm of the operator, defined from the product
\begin{equation}
	\label{EnergyScalarProduct}
\langle u_1,\! u_2\rangle_{_{E}} = \Big\langle\begin{pmatrix}
		\phi_1 \\
		\psi_1
	\end{pmatrix}, \begin{pmatrix}
		\phi_2 \\
		\psi_2
	\end{pmatrix}\Big\rangle_{E}=
	\frac{1}{2} \int_{0}^{1} \left(w(\chi)\bar{\psi}_1 \psi_2 + p(\chi)  \partial_\chi\bar{\phi}_1\partial_\chi\phi_2 + q(\chi)\bar{\phi}_1 \phi_2 \right)  d\chi,
\end{equation}
An equivalent definition is the one which directly relates the level sets of the pseudospectrum to the space of possible new eigenvalue positions after a perturbation,
\begin{equation}\label{pseudopert}
	\sigma^\epsilon(L)=\{\lambda\in\mathbb{C},\exists\delta L,\|\delta L\|<\epsilon:\lambda\in\sigma(L+\delta L)\}.
\end{equation}
Note that this second definition involves any perturbation to $L$ which has a small energy norm, including ones which can potentially be related to a physical modification of the environment of the black hole, but also ones which completely change the nature of the operator (e.g. changing the structure of the derivatives). That said, it was shown in ~\cite{JaramilloPRX} that the instability is in fact triggered by \textit{physical} perturbations, encoded in the addition of a perturbation function $\delta V$ to the effective potential, without disturbing the structure of the differential part of the operator. Additionally, it was shown that the degree to which the spectrum is destabilised depends strongly on the ``high wave-number" content of the perturbation, that is, the sharpness of the variation of $\delta V$ in $r$.

While the particular choices for the perturbations $\delta V$ used in ref.~\cite{JaramilloPRX} may not correspond to the addition of classically reasonable matter content to the system~\cite{Cardoso2024}, they are a proof of principle which shows that whatever the perturbation may be, as long as it has a large enough gradient in $r$, it will trigger the instability. Ref.~\cite{Warnick} in fact explicitly shows the relation between the magnitude of the derivatives of $\delta V$ and the rate of displacement of the QNMs in a specific example, further solidifying this result.

The above-mentioned stability of the fundamental mode was also one of the key results, which can directly be related to the fact that gravitational wave observations of compact object collisions which result in a black hole as an end state appear to contain a part which matches well with a fundamental-mode-dominated ringdown~\cite{LIGO}.

\subsection{Emergence of new long-lived modes}

The second type of ``instability" is due to the emergence of new mode branches. It is important to understand that the characteristics of the spectrum depend strongly on the shape of the potential $V$. For the axial gravitational case, the potential has a single barrier with a peak close to the photon sphere, from which it decreases to zero exponentially (in $r^*$) towards the horizon and polynomially towards infinity. The corresponding ``barrier top" modes are not very long-lived (in terms of the characteristic scale of the problem). However, seemingly small perturbations can lead to a qualitative change in the shape of the potential, such as the addition of a well which goes below the asymptotic values, or of a second barrier (or bump). The former can lead to the presence of bound states, while the latter to slowly decaying ``shape resonances"~\cite{BindelZworski}.

Some examples of such qualitative modifications to the potential in the context of QNMs are the double barrier model in~\cite{Barausse2014}, or some of the models explored in~\cite{Cardoso2024}, such as the addition of a perturbatively small mass parameter. As the new families of modes that these modifications introduce can have a slower decay than the fundamental ``barrier top" QNM, the new fundamental mode and first overtones can be said to have been displaced disproportionately to the size of the perturbation, even if the original modes (which are no longer the fundamental and first overtones) happen to still be present in the new spectrum with only a slight displacement. In other words, what can occur is that the label of ``fundamental" and of the overtone numbers may jump to modes in the new branch (according to the usual assignment of these labels), rather than the old modes being displaced. This can also happen by changing the parameters of a problem such that QNMs in two different branches which are already present can switch roles as the fundamental mode, such as in the overtaking of the fundamental oscillatory mode by a de Sitter mode discussed in~\cite{Destounis2023}.

It is important to note that this by itself is distinct from the usual definition of (perturbative) spectral instability discussed above, in which already existing modes are displaced by large distances in the complex plane. It is also interesting that for a spectrally unstable system such as the case of QNMs, adding, say, a bump to the effective potential, can lead to a combination of both of the above effects: the emergence of new long-lived modes, as well as the large displacement of (some of) the already existing modes. This is precisely the case in the example below, for which the perturbed potential goes from having a single barrier to a double barrier, the latter being akin to the case dubbed a ``well on an island" in ref.~\cite{BindelZworski}.

\section{Potential with a bump}\label{s3}

\begin{figure}
\centering
\includegraphics[scale=.4]{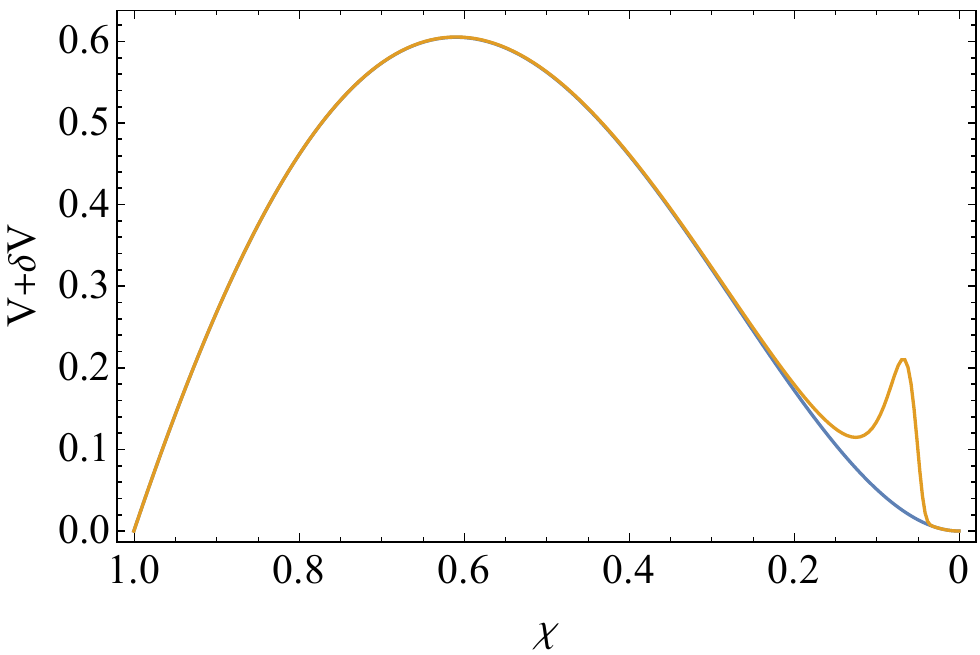}
\caption{Potential for the $\ell=2$ axial gravitational perturbation on Schwarzschild, with an added Gaussian bump at $r_b=30M$. The horizontal axis is the compactified radial coordinate $\chi$, and for illustrative purposes an amplitude $a=0.2$ has been used, which is 40 times larger than the one used for computation.}
\label{f1}
\end{figure}

This section presents an analysis of the above-mentioned perturbation $\delta V$ in the form of a gaussian bump (see fig.~\ref{f1}),
\begin{equation}\label{pert}
\delta V=f(r)\,a\exp(-\frac{(r-r_b)^2}{2s^2}),
\end{equation}
where $a$, $r_b$ and $s$ are positive constants, and the multiplication by $f(r)$ is to ensure that the total potential $V+\delta V$ still has the appropriate tendency to zero when the horizon is approached. Ref.~\cite{elephant} analysed in detail the position of the new long-lived fundamental mode in the presence of such a perturbation. Here we will rather focus on some qualitative features of the modified spectrum as a function of the free parameters in $\delta V$, as well as on quantifying the magnitude of this perturbation using the energy norm. We will use the computational tools employed in ref.~\cite{JaramilloPRX}, namely, a Chebyshev-Lobatto grid in $\chi$ and a pseudospectral approximation to the differential operator $L$ and to the integration operator involved in the energy product.

\subsection{Fundamental mode (in)stability}\label{s31}

\begin{figure}
\centering
\includegraphics[scale=.55]{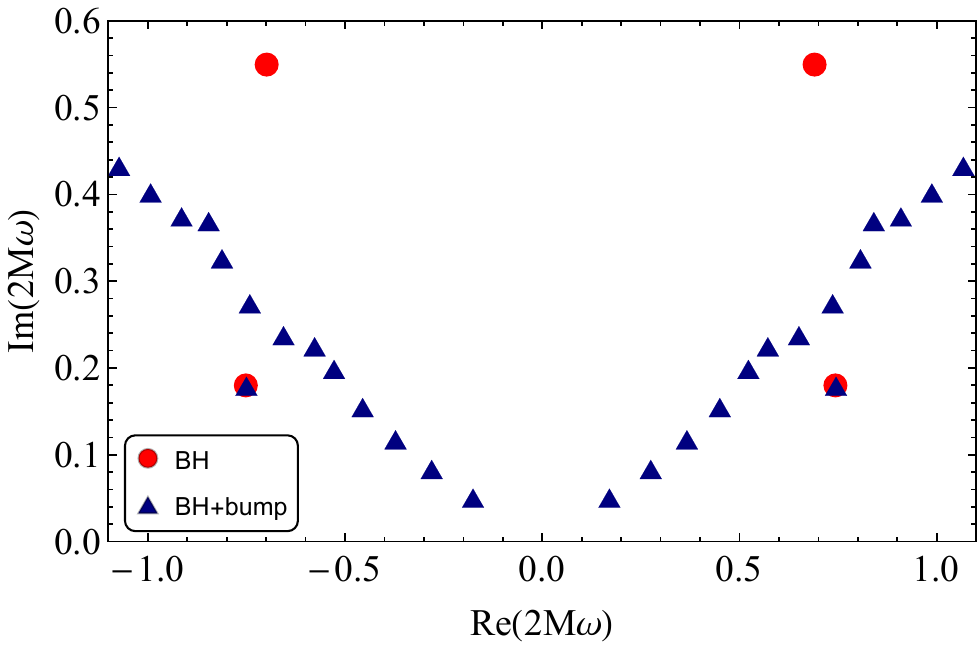}
\caption{QNMs of an axial gravitational $\ell=2$ perturbation of Schwarzschild, with and without a gaussian perturbation. The units are set to $2M=1$. The gaussian bump has parameters $a=0.005$, $r_b=25$, $s=4$, and the spectrum is calculated with $N=400$ grid points. The unperturbed BH fundamental mode is located at $\pm0.74734+0.17792i$, the mode which seemingly overlaps with it after the perturbation is at $\pm0.74729+0.17780i$, and the new fundamental mode is at $\pm0.17287+0.048828i$. The non-convergent ``branch-cut" modes have been removed from the plot (see \cite{JaramilloPRX}).}
\label{f2}
\end{figure}

Let us begin with a particularly illustrative example of a perturbation of this type, which will be the centre-point of this analysis. We set the units to the characteristic scale of the problem by taking $2M=1$, and we set a (seemingly) small amplitude for the bump $a=0.005$, a position for the peak at $r_b=25$ and a width $s=4$. The spectrum of axial $\ell=2$ modes with this perturbation is shown in fig.~\ref{f2}. We see that there is indeed a new branch of modes, some of which decay more slowly than the unperturbed BH fundamental mode. In this sense, the distance between the old and new fundamental mode does indeed seem quite large compared to the size of the perturbation, as discussed in ref.~\cite{elephant}. However, it is also clear form fig.~\ref{f2} that the spectrum after the perturbation contains a mode which coincides with the unperturbed fundamental one (in fact it is only $\sim10^{-3}$ away), implying that this mode was actually stable under the perturbation.

This is therefore a case in which it is the qualitative change in the shape of the potential has lead to the appearance of new long-lived modes, while \emph{part of} the old spectrum has remained stable, in this case only the BH fundamental mode. From the first overtone onwards, the BH spectrum is in fact destabilised, much like it is in some of the cases studied in ref.~\cite{JaramilloPRX}.

The behaviour of the new fundamental mode depends strongly on the parameters of the gaussian bump $a$, $r_b$ and $s$, and while a detailed analysis of this dependence is not within the scope of this work (see \cite{elephant,Cardoso2024} for a quantitative analysis of part of the parameter space), we will make some general remarks regarding the behaviour we have observed from a few spectra.
\begin{itemize}
	\item Increasing the amplitude $a$ tends to decrease the imaginary part of the fundamental mode, as the modes trapped between the two peaks (the bump and the light-ring peak) need to tunnel out of a larger barrier to decay. Conversely, if $a$ is made smaller, the imaginary part increases. At around $a=10^{-5}$ the longest-lived of these new modes is no longer the fundamental one, as its imaginary part is larger than that of the BH mode.
	\item Increasing the radial position of the bump $r_b$ decreases the imaginary part of the new modes, as well as destabilising the old BH spectrum more strongly. We will make some remarks regarding the reason for this in the next section.
	\item Increasing the width of the bump $s$ also makes the mode longer-lived, since this increases the tunnelling (Agmon) distance~\cite{BindelZworski}. On the other hand, a larger $s$ (at a fixed energy norm) makes the old BH spectrum more stable, since then the perturbation has a lesser ``high-wavenumber" content, as discussed in~\cite{JaramilloPRX}.
\end{itemize}

It is also worth noting that while we identify these longer-lived modes as a new branch due to the qualitative change in the potential and the stability of the BH fundamental mode, from the numerical results alone it is not clear where exactly this new branch becomes entwined with the perturbed BH overtones. To identify which modes go to infinity and which to BH overtones in the zero perturbation limit, a more detailed study which traces the migration of individual overtones would be required.

\subsection{Flea or elephant?}\label{s32}

\begin{figure}
\centering
\includegraphics[scale=.45]{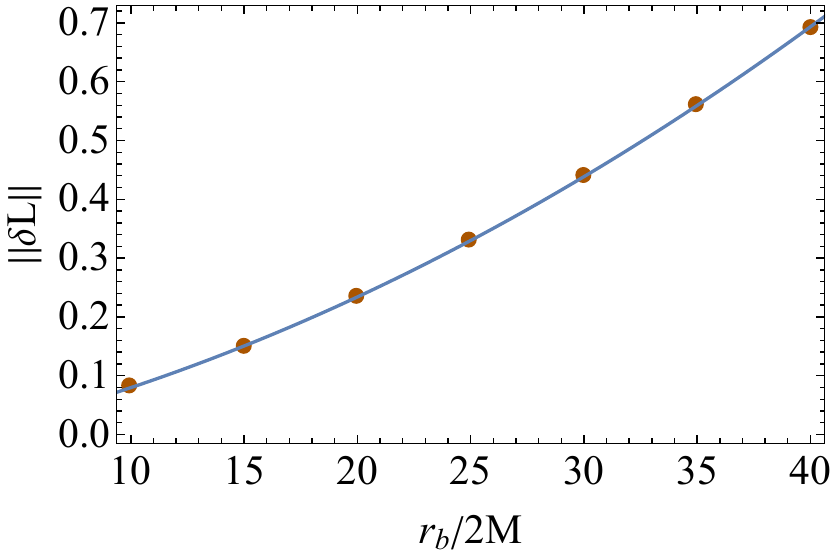}
\hspace*{3mm}
\includegraphics[scale=.45]{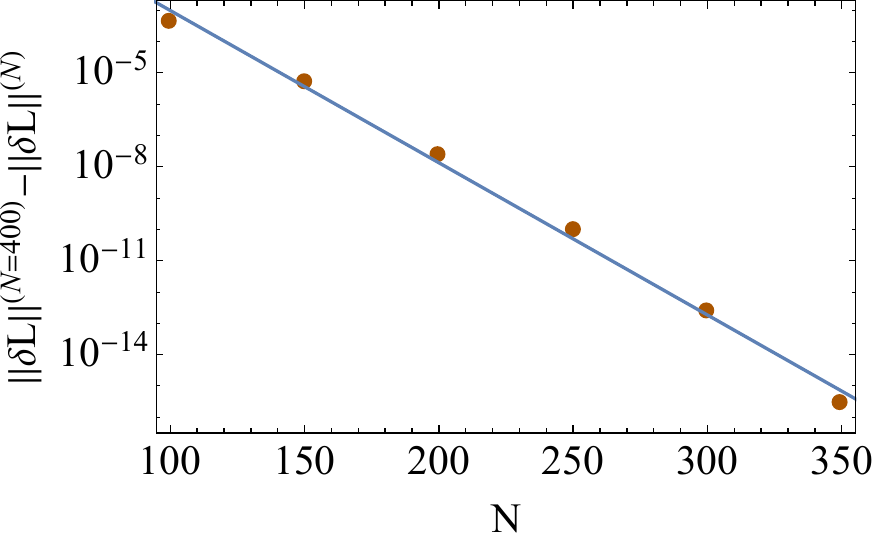}
\caption{Left: energy norm of $\delta L$ as function of $r_b$, for $s=4$ and $a=0.005$, in units $2M=1$. The quadratic fit is $-0.024 + 0.0079 r_b + 0.00025 r_b^2$ (the variance is $\sim10^{-6}$, though at smaller radii this relation can be expected to start breaking down, since the norm must be positive). Right: energy norm of $\delta L$ calculated with $N$ points, dubbed $\|\delta L\|^{(N)}$, subtracted from a reference value $\|\delta L\|^{(400)}$, for the same case with $r_b=25$. The vertical axis is log scaled to showcase the exponential convergence (the slight discrepancy from the linear fit is due to the finite $N$ reference value).}
\label{f3}
\end{figure}

The perturbation operator being added to $L$ can be written as
\begin{equation}
	\delta L=\begin{pmatrix}
		0&0\\ \frac{\delta q}{w}\mathbb{I}&0
	\end{pmatrix},
\end{equation}
where $\delta q=|g'|\delta V$. In order to give a physical measure of the size of this perturbation, the energy norm of $\delta L$ can be computed. Contrary to what might be expected from the small $a$ parameter, the energy norm of the example case used for fig.~\ref{f2} is actually quite significant: in units of the horizon scale, it is approximately $0.33$, on the very high end of what can reasonably be considered a ``perturbation". The reason for this apparent discrepancy between the intended smallness in the choice of $a$ and the large energetic contribution of this perturbation lies in the simple fact that the energy measure comes from an integral related to the full 3 dimensional space of constant time slices~\cite{Gasperin2021}, rather than just the 1 dimension of the wave problem (although in the end it simplifies to the latter). It therefore encodes the fact that a perturbation at a large radius would require a thick shell of this same radius, the size and matter content of which would scale with $r^2$. This is indeed the scaling we can observe in the left plot of fig.~\ref{f3}, where the energy norm is calculated as a function of $r_b$ (with all other parameters remaining the same), and fitted to a parabola.

Therefore, the increased destabilisation of the BH QNMs (effectively, the lowering of the mode branches seen in fig.~\ref{f2}) for a larger $r_b$ which was commented above can be related to precisely this increase of the energy norm. The dependence of this energy on the parameters $a$ and $s$ is just as predictable: an increase in both these parameters leads to a proportional (linear) increase in the energy norm.

\subsection{Size vs instability}\label{s33}

This example has shown the importance of quantifying the size of perturbations added to the problem, since, for instance, the increase of the energy contained in perturbations at larger radii is something that could easily have been overlooked otherwise. However, using the energy norm in particular, while having many advantages~\cite{Gasperin2021}, may not be the most adequate choice in some respects. One particular issue, raised in~\cite{Boyanov2023}, is the fact that the energy norm of the resolvent operator, used to calculate the pseudospectrum, is not well behaved in a large part of the complex plane, which includes the vicinity of most (if not all) QNMs. Numerically, this norm tends to a divergence in the limit of infinite grid point number $N$ in most of the upper half of the complex plane. Since the issue in that case stems from the presence of additional eigenmodes of a lower regularity class~\cite{Warnick2013}, and not simply from a numerical problem, it is likely to be a generic property of other setups as well.

One may then ask whether this issue extends to calculating the energy norm of other operators as well, particularly that of $\delta L$, since the pseudospectrum can equivalently be defined from its norm (albeit for a very large set of perturbations). Fortunately, it appears that the norm of this operator actually does have a good convergent behaviour. The second plot in fig.~\ref{f3} shows a convergence test in a representative example. The result is clear: the convergence is in fact exponential. Such convergence was previously observed for other quantities computed with this discretisation scheme, such as the spectrum itself (see fig.~8 of~\cite{JaramilloPRX}), but had thus far not been tested for energy norms of operators, except for the case of the non-convergent resolvent norm in~\cite{Boyanov2023}.

The convergence of $\|\delta L\|$ in fact confirms that the issue with the resolvent norm studied in~\cite{Boyanov2023} goes beyond the particular numerical implementation. It also gives an appealing potential alternative approach to calculating the pseudospectrum by exploring a sufficiently large space of perturbation operators and applying eq.~\eqref{pseudopert} \cite{Trefethen2005,Arean}. However, there would be two issues with such an approach. First, it would be computationally very expensive to attempt to span a ``full" space of perturbation operators $\delta L$. This would not be a critical impediment, at least for a small numerical resolution $N$. However, the second and most crucial issue is the fact that the result would differ depending on the resolution, as a higher $N$ could capture perturbations with a higher wave-number content, which would destabilise the spectrum ever more strongly. It is not clear that the limit of operators $\delta L$ with the same energy norm but with ever higher gradients in $r$ (which would need a correspondingly higher $N$ to be resolved) would lead to a convergent definition of the pseudospectrum, or if this issue would turn out to be equivalent to the non-convergence observed in the resolvent approach.

Testing whether this claim is true, while absolutely crucial, goes beyond the scope of the present work. If it were indeed proven true, then a consistent definition of a QNM pseudospectrum would require a modification of the scheme summarised above. One example of such a modification would be the use of norms with higher order spatial derivatives, as introduced in~\cite{Warnick2013}, and applied to the pseudospectral calculation in~\cite{Boyanov2023} (see also~\cite{Besson}). However, a reasonable physical interpretation of such norms and their associated stability analyses would need to be devised.

\section{Discussion}\label{s4}

The spectral instability of BH QNMs is by now a well established result in the field of black hole spectroscopy. As we have seen here, QNMs are susceptible to (al least) two different types of instability: either the direct migration of the already existing QNMs by a large distance in the complex plane (``perturbative" instability), or the appearance of new branches of modes to which the new label of fundamental or overtone number are assigned, and which are far away from their unperturbed counterparts (``branch" instability).

One important aspect in analysing both perturbative and branch instabilities is quantifying the size of the perturbations introduced into the system. A physically reasonable measure of this size is given by the energy norm~\cite{Gasperin2021}, which comes from an inner product space associated to the energy of the linear field. As we have seen in the above examples, a seemingly small perturbation to the effective potential can in fact have a large energy norm, and have a correspondingly large destabilising effect on the spectrum.

Keeping track of this norm is therefore crucial. Indeed, in the example of a gaussian bump studied in the present work, there is a clear correlation between the energy norm of $\delta L$ and the distance between the old and new fundamental modes. However, there are two issues with establishing a direct one-to-one relation between this norm and the expected degree of destabilisation of the spectrum. First, the fact remains that (some of) the original BH modes can in fact remain stable in spite of the appearance of the new branches of longer-lived modes. Second, the degree to which these original modes are actually destabilised does not depend only on the energy norm, but also on the high-wave-number content of the perturbation involved, as observed in~\cite{JaramilloPRX}. For the gaussian bump perturbation used here, decreasing the width of the bump decreases its associated energy norm, but the resulting sharper variation in $r$ can in fact lead to an increase in the instability of the original spectrum. Exploring this issue in detail is particularly difficult because of the numerics involved, since a sharper bump requires a higher resolution to be captured, making the degree to which a bump of any given energy norm can destabilise the spectrum hard to establish.

This difficulty can in fact be seen as a potential issue with providing a convergent result for the pseudospectrum in the energy norm, since if there were such a result, a bound on the possible migration of modes could be easily placed through eq.~\eqref{pseudopert}. However, obtaining such a convergent pseudospectrum has been an elusive task, as discussed in~\cite{Boyanov2023}. Finding a solution to this issue would likely require changing parts of the above-described prescription to this calculation, as is currently being explored by the present author and collaborators~\cite{Besson}.

Regarding the observability of these instabilities in gravitational wave signals, the results of refs.~\cite{Jaramillo2021,Spieksma2024} suggest that while environmental perturbations are detectable in time-domain evolution, their effect on ringdown signals is not as disproportionately large as it is on the QNM spectrum itself. However, a systematic study of the effect of different types of perturbations, particularly involving the branch instability analysed here, is lacking.

Overall, the study of the QNM spectral instability has led to a myriad of different results in many different spacetime setups, but there are just as many open questions left to be addressed in the coming years.

\section*{Acknowledgements}

The author would like to thank Jose Luis Jaramillo for a detailed discussion and feedback on this manuscript, as well as Vitor Cardoso, Kyriakos Destounis and Rodrigo Panosso Macedo for our many group discussions on this topic. Financial support was provided by the European Union’s H2020 ERC Advanced Grant “Black holes: gravitational engines of discovery” grant agreement no. Gravitas–101052587, and by the Spanish Government through the Grants No. PID2020-118159GB-C43, PID2020-118159GB-C44, PID2023-149018NB-C43 and PID2023-149018NB-C44 (funded by MCIN/AEI/10.13039/501100011033).

\nocite{*}
\bibliography{Bibliografia}
\bibliographystyle{ieeetr}
\end{document}